\documentclass[fleqn,twoside]{article}
\usepackage{espcrc2}
\usepackage{graphicx,amsmath}

\newcommand{\fcaption}{%
\vspace*{-1.0cm}
\caption%
}
\title{The Structure of Projected Center Vortices at Zero and Finite Temperature}

\author{R. Bertle$^{\mathrm{a}}$%
	\thanks{Talk presented by R. Bertle. Supported in part by FWF P11387-PHY.}
        ,
        M. Faber\address{Inst. f\"ur Kernphysik, Technische Universit\"at Wien\\
        A--1040 Vienna, Austria}
	,
	J. Greensite\address{Physics and Astronomy Dept., San Francisco State University\\
        San Francisco, CA~94117, USA}
	and
	{\v S}.  Olejn\'{\i}k\address{Institute of Physics, Slovak Academy of Sciences\\
        SK--842 28 Bratislava, Slovakia}
}
	
\begin{document}

\begin{abstract}
We investigate the structure of center projected vortices of SU(2) lattice gauge theory at zero and finite temperature.
At zero temperature we find, in agreement with the area law behaviour of Wilson loops, that most of the P-vortex plaquettes are parts of a single huge vortex.
This vortex is an unorientable surface and has a very irregular structure with many handles.
Small P-vortices, and short-range fluctuations of the large vortex surface, do not contribute to the string tension.
At finite temperature P-vortices exist also in the deconfined phase.
However, they form cylindric objects which extend in time direction and consist only of space-space plaquettes.
\end{abstract}

\maketitle

\section{INTRODUCTION}

Using the direct version of maximal center gauge we identify projected (P-)vortices by center projection \cite{DFGGO98}.
We map the SU(2) link variables $U_\mu(x)$ to $Z_2$ elements $Z_\mu(x) = \text{sign Tr} [ U_\mu(x) ]$.
The plaquettes with $Z_{\mu\nu}(x) = Z_\mu(x) Z_\nu(x+\hat{\mu}) Z_\mu(x+\hat{\nu}) Z_\nu(x) = -1$ we call ``P-plaquettes.''
The corresponding dual plaquettes form a closed surface in 4 dimensions.

\section{ZERO TEMPERATURE}

\subsection{String tension and smoothing}
The distribution of P-vortices in space-time determines the string tension $\sigma$ in center projection which agrees very well with $\sigma$ from full Wilson loops \cite{DFGGO98}.
If $p$ is the probability that a plaquette belongs to a P-vortex, we get for the expectation value of a Wilson loop of size $A = I \times J$ assuming the independence of piercings of the loop
\begin{eqnarray}
\langle W_{cp}(I,J) \rangle & = & \left [ (1-p)1 + p(-1) \right ]^A \\
& = & (1-2p)^A = e^{- \sigma_{cp} A} \approx e^{-2pA} \,,\nonumber 
\end{eqnarray}
where the string tension in center projection is
\begin{equation}\label{sigmatop} 
\sigma_{cp} = - \text{ln} (1-2p) \approx 2p \,.
\end{equation}
\begin{figure}
\includegraphics[width=\linewidth]{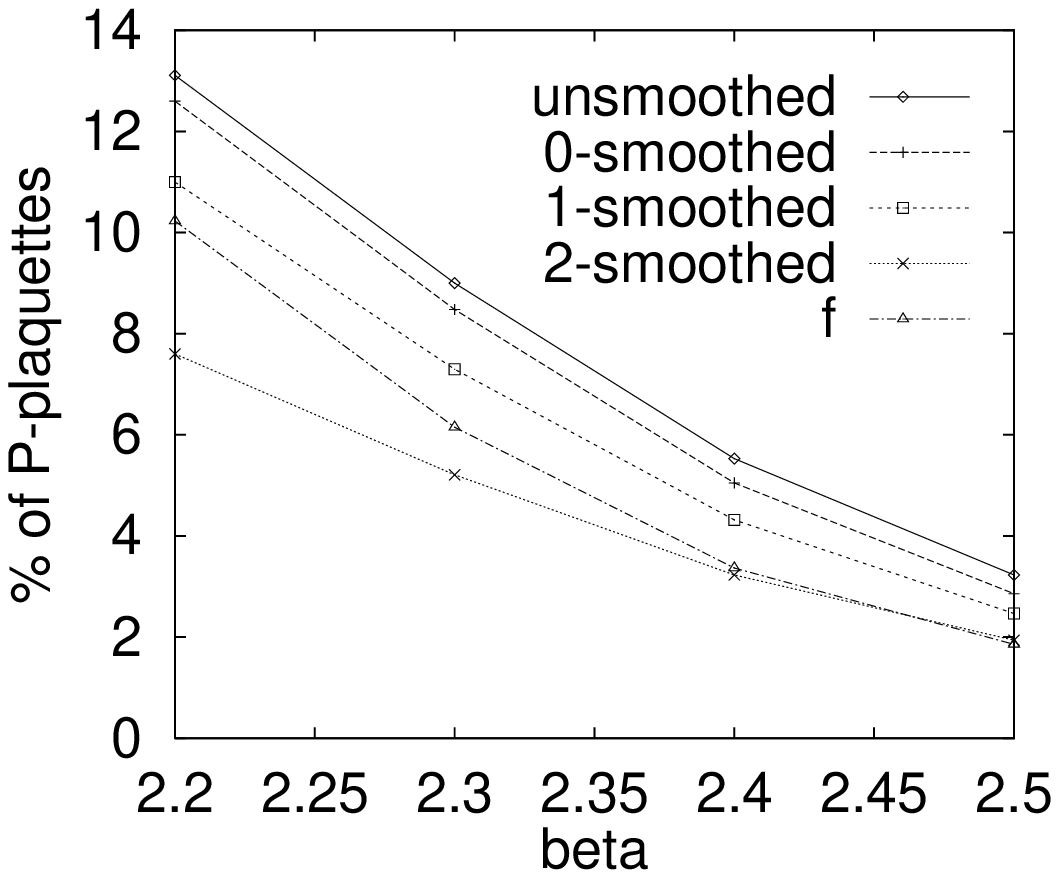}%
\fcaption{Percentage $p$ of P-vortex plaquettes.}
\label{nplneg}
\end{figure}
$p$ scales nicely with the inverse coupling $\beta$ \cite{DFGGO98,LRT98}.
However, for small vortices the independence assumption of piercings is not fulfilled, simply because one piercing is always correlated with another piercing nearby.
Hence, small vortices do not contribute to the string tension nor do small fluctuations of the P-vortex surface.
This can be seen from Fig.~\ref{nplneg}:
$p$ -- labeled with ``unsmoothed'' -- scales, but it is higher than the fraction of P-plaquettes $f$ infered from the measured string tension $\sigma$ using $f = (1-e^{-\sigma})/2$ and assuming independent piercings.

\begin{figure}
\includegraphics[width=0.8\linewidth]{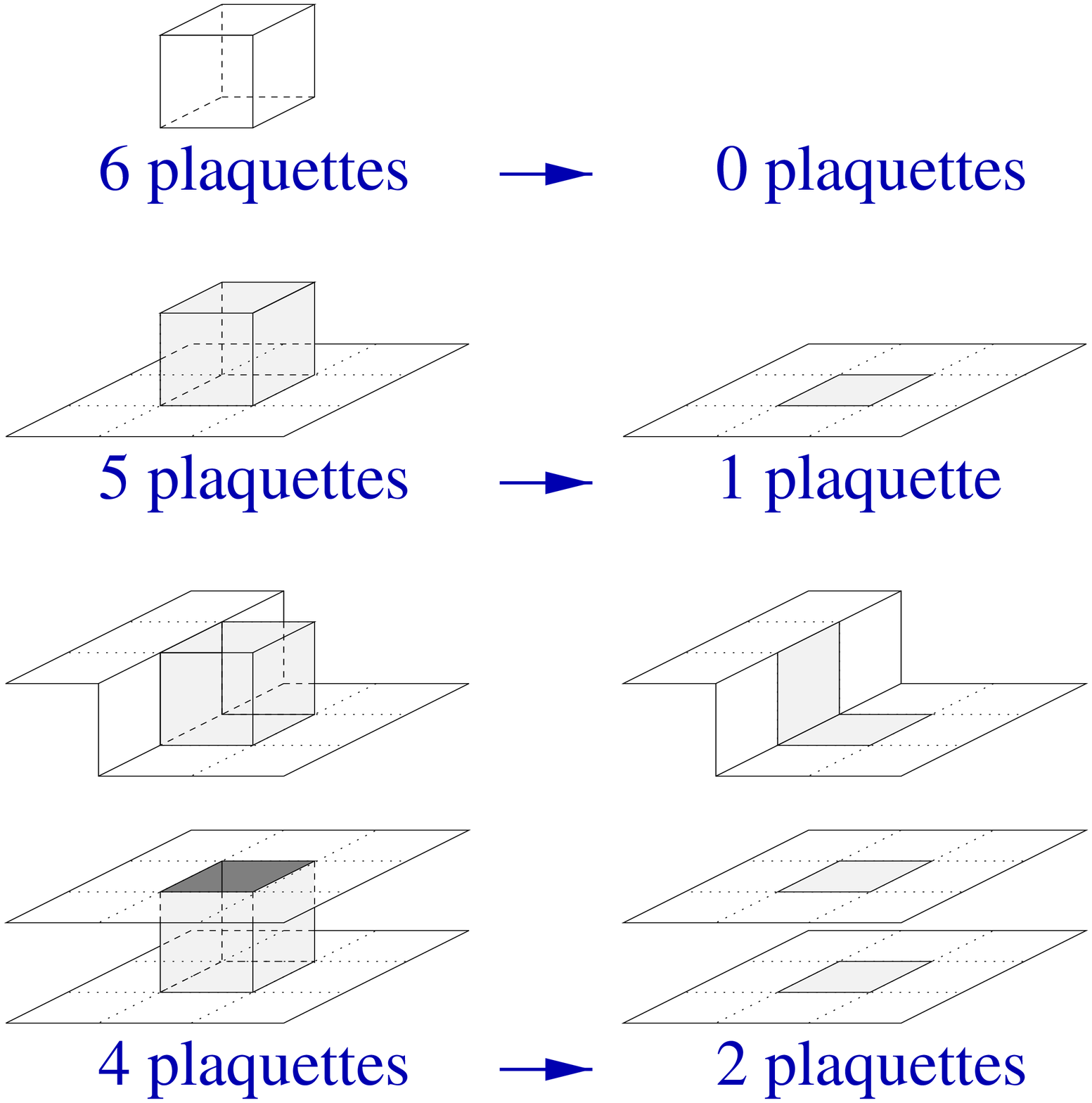}%
\fcaption{Various smoothing steps for P-vortices.}
\label{smooth}
\end{figure}
To understand the discrepancy between $p$ and $f$ in more detail we remove short range fluctuations which are unimportant for $\sigma$. We introduce several smoothing steps which are depicted in Fig.~\ref{smooth}.
First, scanning through the lattice we iteratively identify cubes with 6 P-plaquettes and remove them which is called 0-smoothing.
In the next steps we substitute cubes with 5 or 4 P-plaquettes by cubes with 1 resp.\ 2 complementary P-plaquettes which is called 1- resp.\ 2-smoothing.

\begin{figure}[!b]
\includegraphics[width=\linewidth]{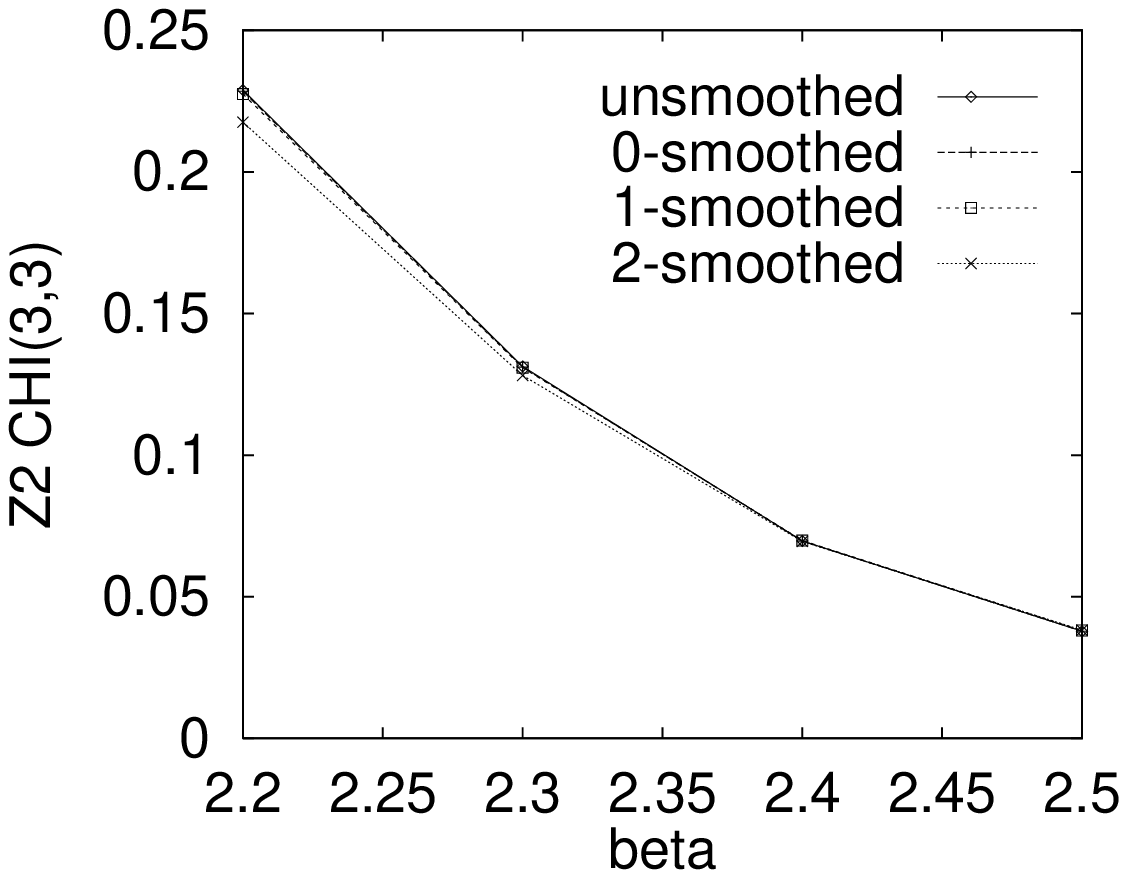}%
\fcaption{Creutz ratios for different smoothing steps.}
\label{creutz}
\end{figure}
As can be seen from Fig.~\ref{nplneg}, the value of $p$ nicely approaches $f$ with increasing smoothing step, especially for larger values of $\beta$ where P-plaquettes get less dense.
Further we check the Creutz ratios extracted from P-configurations after various smoothing steps.
It is clearly seen that 0- and 1-smoothing do not change the Creutz ratios, only 2-smoothing shows a deviation of $\sim 5\%$ for small $\beta$ (Fig.~\ref{creutz}).

\subsection{Topology}
P-vortices have to percolate in order to give the Wilson loop an area law behaviour resulting in a finite string tension.
Thus we check the size of P-vortices.
The result is that around 90\% of all P-plaquettes are part of one huge P-vortex.
All other P-vortices are rather small and should not contribute to $\sigma$.
After 2-smoothing this huge vortex contains almost all (over 99\%) of all P-plaquettes.

\begin{figure}[!ht]
\includegraphics[width=\linewidth]{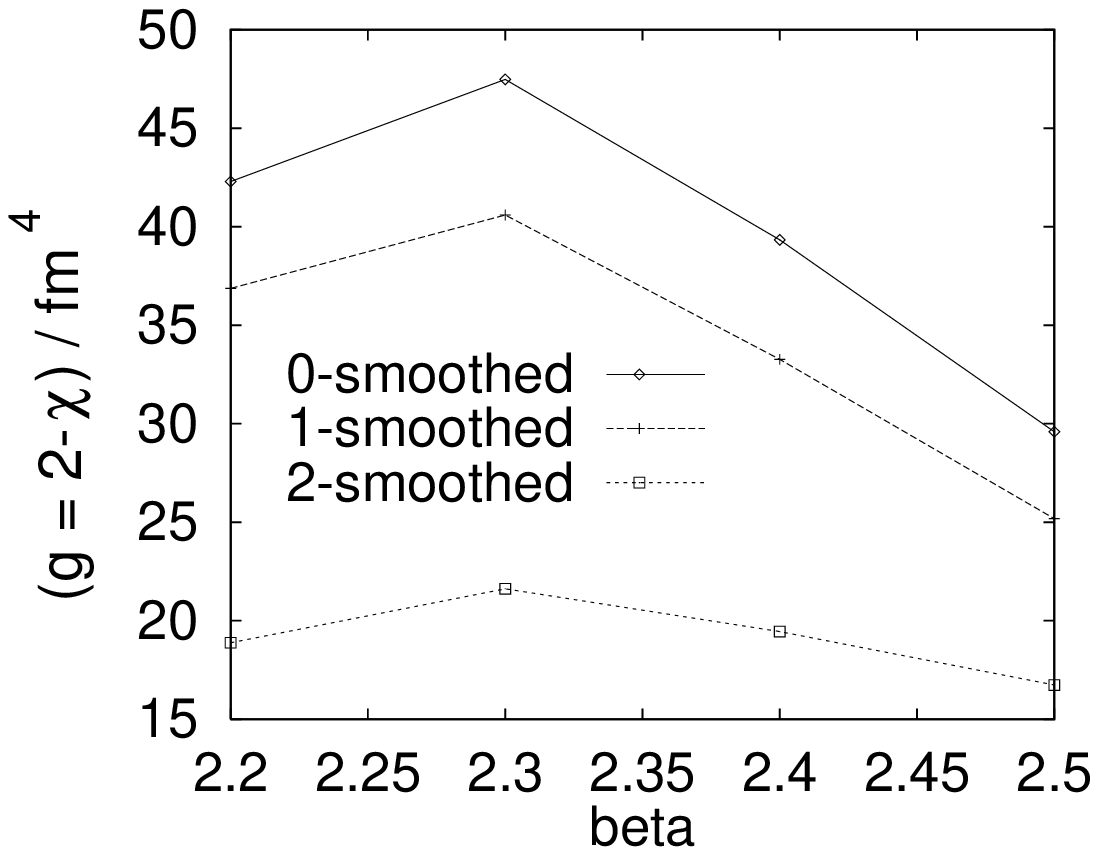}%
\fcaption{Genus $g$ of P-vortices per fm$^4$.}
\label{euler}
\end{figure}
Next we calculate the type of homomorphy of the surface of the dominating P-vortex.
It is determined by $a$) the orientation behaviour, $b$) the Euler characteristic $\chi$.

The simulation shows that without exception large vortices are unorientable even after smoothing; apparently the smoothing procedure does not remove all of the local structures (e.g. ``cross-caps'') responsible for the global non-orientability.

The Euler characteristic $\chi$ is defined as $\chi = {\cal N}_0 - {\cal N}_1 + {\cal N}_2$ where ${\cal N}_i$ is the number of vertices, links and plaquettes resp.
$\chi$ is related to the genus $g$ by $\chi = 2 - g$; an unorientable surface of genus $g$ is homeomorphic to a sphere with $g$ attached ``cross-caps''.
Fig.~\ref{euler} shows that after 2-smoothing $g$ roughly scales. This is not compatible with a self-similar short-range structure below the confinement length scale i.e.\ a fractal structure.
\section{FINITE TEMPERATURE}

In this section we will extend our study of P-vortex topology, and the effect of our smoothing steps on P-vortices, to the finite temperature case; see also \cite{LTER99}.
In the deconfinement phase there is a strong asymmetry of P-plaquette distributions.
The density of space-time plaquettes decreases explaining why the string tension of timelike Wilson loops is lost and $\sigma$ of spatial loops is preserved \cite{KLL95}.
The dominance of the largest P-vortex is weaker in deconfinement, but as expected it is still there.

\begin{figure}[!h]
\includegraphics[width=\linewidth]{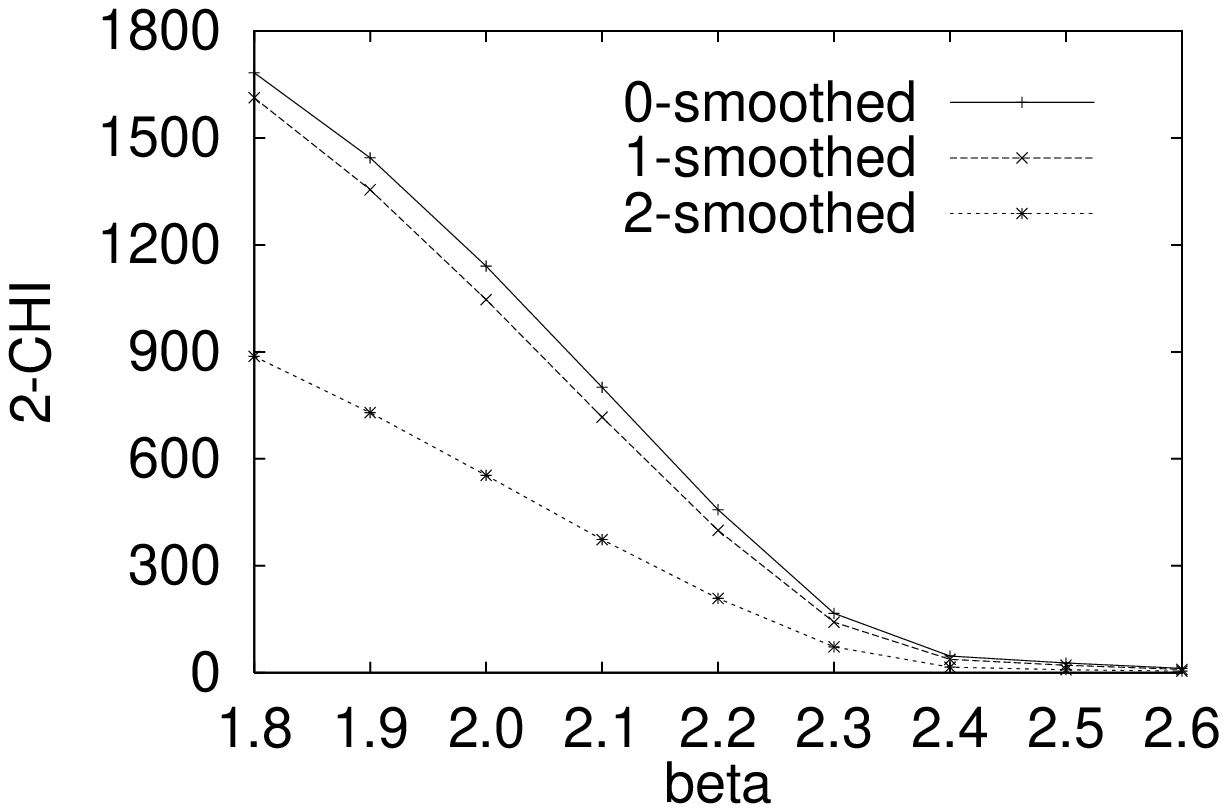}%
\fcaption{Genus of the dominating vortex.}
\label{euler4}
\end{figure}
\begin{figure}[!t]
\includegraphics[width=0.7\linewidth]{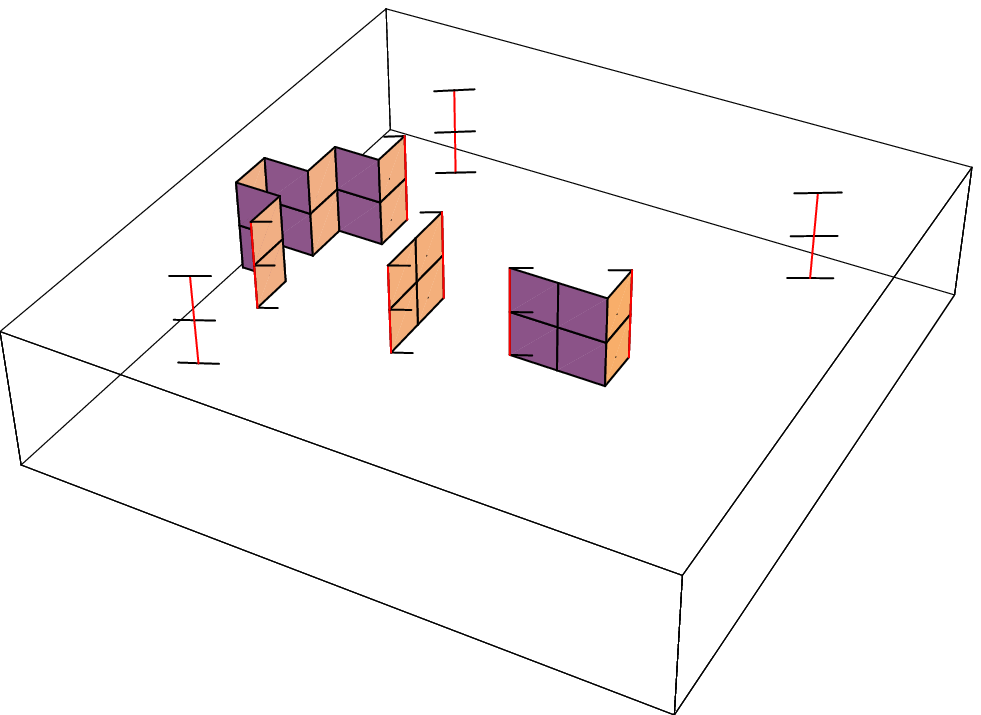}%
\fcaption{Dual P-plaquettes in a z-slice for a typical field configuration.%
}
\label{math4}
\end{figure}

In contrast to the zero temperature case P-vortices get orientable in the deconfinement phase.
The dual P-plaquettes form cylinders in time direction, closed via the periodicity of the lattice.
For high temperatures the Euler characteristic $\chi$ approaches $0$ as shown in Fig.~\ref{euler4}.
The largest P-vortex has the topology of a torus.
Fig.~\ref{math4} shows a cut through a typical field configuration at $\beta=2.6$ on a $2\cdot 12^3$-lattice.

\section{CONCLUSIONS}

We have investigated the size and topology of P-vortices in SU(2) lattice gauge theory; P-vortices are surfaces on the dual lattice which 
lie at or near the middle of thick center vortices.
In the confined phase the four-dimensional lattice is penetrated by a single huge P-vortex of very complicated topology.
It is unorientable and has many handles.
There exist also a few very small vortices.
These and short range fluctuations of the large P-vortex don't contribute to the string tension.
Keeping the Creutz ratios constant, we could remove those fluctuations by a smoothing procedure.

In the deconfined phase, we found a strong space-time asymmetry.
P-vortices at finite temperature are mainly 
composed of space-space plaquettes forming time-like surfaces 
on the dual lattice.
They are orientable, closed via the periodicity of
the lattice in the time direction and have the topology of a torus.
The dominance of the largest vortex is not as strong as in the 
zero temperature case.

Further details and an expanded discussion can be found in \cite{BFGO99}.
\bibliography{bertle}
\end{document}